\documentclass[notitlepage,nofootinbib,preprintnumbers,amssymb,superscriptaddress]{revtex4-2}
\usepackage{amsfonts,amssymb,mathtools,graphicx,color,bm}
\definecolor{ultramarine}{rgb}{0.07, 0.04, 0.56}
\definecolor{cadmiumgreen}{rgb}{0.0, 0.42, 0.24}
\definecolor{indigo(dye)}{rgb}{0.0, 0.25, 0.42}
\usepackage[linktocpage=true]{hyperref}
\hypersetup{
colorlinks=true,
citecolor=ultramarine,
linkcolor=cadmiumgreen,
urlcolor=indigo(dye),
}

\usepackage{autobreak}
\newcommand{\D}{{\rm d}}
\newcommand{\DX}[1]{{\cal X}_{#1}}
\newcommand{\DXup}[1]{{\cal X}^{#1}}
\newcommand{\dn}{\nu}
\newcommand{\aH}{\ti{\alpha}}
\newcommand{\MD}{M_{\rm D}}
\newcommand{\fr}[2]{\frac{#1}{#2}}
\newcommand{\pa}{\partial}
\newcommand{\ti}{\tilde}
\newcommand{\na}{\nabla}
\newcommand{\bra}[1]{\left( #1 \right)}  
\newcommand{\brb}[1]{\left[ #1 \right]}  
  
\newcommand{\be}{\begin{equation}}  
\newcommand{\ee}{\end{equation}}
\newcommand{\bem}{\begin{bmatrix}}
\newcommand{\eem}{\end{bmatrix}}

\newcommand{\ga}{\gamma}

\newcommand{\la}{\lambda}

\newcommand{\mn}{{\mu \nu}}
\newcommand{\mB}{\mathcal{B}}
\newcommand{\mC}{\mathcal{C}}
\newcommand{\mD}{\mathcal{D}}
\newcommand{\mE}{\mathcal{E}}
\newcommand{\mF}{\mathcal{F}}
\newcommand{\mG}{\mathcal{G}}

\newcommand{\mJ}{\mathcal{J}}
\newcommand{\mK}{\mathcal{K}}
\newcommand{\mL}{\mathcal{L}}
\newcommand{\mM}{\mathcal{M}}

\newcommand{\mO}{\mathcal{O}}

\newcommand{\mW}{\mathcal{W}}
\newcommand{\mZ}{\mathcal{Z}}
\newcommand{\Rs}{{}^{(3)}\!R}
\newcommand{\bRs}{{}^{(3)}\!\mybar{R}}

\makeatletter
\newsavebox\myboxA
\newsavebox\myboxB
\newlength\mylenA
\newcommand*\mybar[2][0.75]{%
    \sbox{\myboxA}{$\m@th#2$}%
    \setbox\myboxB\null
    \ht\myboxB=\ht\myboxA%
    \dp\myboxB=\dp\myboxA%
    \wd\myboxB=#1\wd\myboxA
    \sbox\myboxB{$\m@th\overline{\copy\myboxB}$}
    \setlength\mylenA{\the\wd\myboxA}
    \addtolength\mylenA{-\the\wd\myboxB}%
    \ifdim\wd\myboxB<\wd\myboxA%
       \rlap{\hskip 0.7\mylenA\usebox\myboxB}{\usebox\myboxA}%
    \else
        \hskip -0.5\mylenA\rlap{\usebox\myboxA}{\hskip 0.5\mylenA\usebox\myboxB}%
    \fi}
\makeatother

\begin{document}

\preprint{YITP-23-46}

\title{Effective description of generalized disformal theories}

\author{Kazufumi Takahashi}
\affiliation{Center for Gravitational Physics and Quantum Information, Yukawa Institute for Theoretical Physics, Kyoto University, 606-8502, Kyoto, Japan}

\author{Masato Minamitsuji}
\affiliation{Centro de Astrof\'{\i}sica e Gravita\c c\~ao  - CENTRA, Departamento de F\'{\i}sica, Instituto Superior T\'ecnico - IST,
Universidade de Lisboa - UL, Av.~Rovisco Pais 1, 1049-001 Lisboa, Portugal}

\author{Hayato Motohashi}
\affiliation{Division of Liberal Arts, Kogakuin University, 2665-1 Nakano-machi, Hachioji, Tokyo 192-0015, Japan}

\begin{abstract}
Generalized disformal transformations enable us to construct the generalized disformal Horndeski theories, which form the most general class of ghost-free scalar-tensor theories to this date.
We extend the effective field theory (EFT) of cosmological perturbations to incorporate these generalized disformal Horndeski theories.
The main difference from the conventional EFT is that our extended EFT involves operators with higher spatial derivatives of the lapse function.
Our EFT also accommodates the generalized disformal transformation of U-DHOST theories.
\end{abstract}

\maketitle

\section{Introduction}\label{sec:intro}

Scalar-tensor theories have been extensively studied as one of the simplest extensions of general relativity, i.e., modified gravity theories.
Although simple, they are useful to explore universal features of generic modified gravity models which contain additional degrees of freedom (DOFs) on top of the spacetime metric.
Moreover, they often exhibit peculiar phenomena that can be observationally tested against general relativity (see Refs.~\cite{Koyama:2015vza,Ferreira:2019xrr,Arai:2022ilw} for reviews).
Among various scalar-tensor theories, the so-called Horndeski theory is the fundamental element, as it is the most general scalar-tensor theory with second-order Euler-Lagrange equations~\cite{Horndeski:1974wa,Deffayet:2011gz,Kobayashi:2011nu}.
Actually, this theory contains many traditional scalar-tensor theories (e.g., the Brans-Dicke theory~\cite{Brans:1961sx}, $f(R)$ theories~\cite{Sotiriou:2008rp,DeFelice:2010aj}, scalar-Gauss-Bonnet theories~\cite{Koivisto:2006xf,Sotiriou:2013qea,Sotiriou:2014pfa,Antoniou:2017hxj,Silva:2017uqg}, and k-essence theories~\cite{ArmendarizPicon:1999rj}) as its particular cases, and in this sense it can be viewed as a class of scalar-tensor theories rather than a single theory.
It should be noted that the Horndeski class is free of the problem of Ostrogradsky ghosts~\cite{Woodard:2015zca,Motohashi:2020psc} thanks to the second-order nature of the Euler-Lagrange equations.
One can further enlarge the space of ghost-free scalar-tensor theories by allowing the Euler-Lagrange equations to contain higher derivatives and imposing the degeneracy condition~\cite{Motohashi:2014opa,Langlois:2015cwa,Motohashi:2016ftl,Klein:2016aiq,Motohashi:2017eya,Motohashi:2018pxg}.
The authors of Refs.~\cite{Langlois:2015cwa,Crisostomi:2016czh,BenAchour:2016fzp} constructed a general class of theories that satisfy the degeneracy condition, which is now called the degenerate higher-order scalar-tensor (DHOST) class.\footnote{DHOST theories literally refer to any higher-order scalar-tensor theories satisfying the degeneracy condition, but they often refer to only those constructed in Refs.~\cite{Langlois:2015cwa,Crisostomi:2016czh,BenAchour:2016fzp}. We also follow this conventional terminology in the present paper.}
A yet further extension of ghost-free scalar-tensor theories is the class of U-DHOST theories~\cite{DeFelice:2018mkq,DeFelice:2021hps}.
Under the unitary gauge, U-DHOST theories are equivalent to spatially covariant gravity theories~\cite{Gao:2014soa,Gao:2018znj,Gao:2019lpz,Motohashi:2020wxj}, which satisfy the degeneracy condition.
Away from the unitary gauge, U-DHOST theories have an apparent Ostrogradsky mode in general, but the mode actually satisfies an elliptic differential equation on a spacelike hypersurface and hence does not propagate.
This instantaneous mode is often called a shadowy mode~\cite{DeFelice:2018mkq,DeFelice:2021hps}.

While imposing the degeneracy condition offers a systematic construction of ghost-free scalar-tensor theories beyond the Horndeski class, it is known that most subclasses of DHOST theories are not phenomenologically viable:
Cosmological perturbations exhibit ghost/gradient instabilities or otherwise tensor perturbations are nondynamical~\cite{Langlois:2017mxy}.
The only viable subclass is the one generated from the Horndeski class via the invertible disformal transformation~\cite{Bekenstein:1992pj,Bruneton:2007si,Bettoni:2013diz},
    \be
    g_\mn\quad\to\quad
    \bar{g}_\mn[g,\phi] = f_0(\phi,X) g_\mn + f_1(\phi,X)\na_\mu\phi\na_\nu\phi,
    \label{conv_disformal}
    \ee
where $\phi$ is the scalar field, $\nabla_\mu$ represents the covariant derivative with respect to the metric~$g_{\mu\nu}$, and $f_0$ and $f_1$ are functions of $(\phi,X)$ with $X\coloneqq g^{\alpha\beta}\pa_\alpha\phi\pa_\beta\phi$.
Note that the invertibility, i.e., the (local) unique existence of the inverse transformation, is guaranteed if the coefficient functions satisfy the following conditions:
    \be
    f_0\ne 0, \qquad
    f_0+Xf_1\ne 0, \qquad
    f_0-Xf_{0X}-X^2f_{1X}\ne 0,
    \ee
with a subscript~$X$ denoting the derivative with respect to $X$.
Such an invertible transformation maps a ghost-free scalar-tensor theory to another ghost-free scalar-tensor theory, keeping the number of dynamical DOFs~\cite{Domenech:2015tca,Takahashi:2017zgr}.
We shall refer to the subclass of DHOST theories that are generated by invertible disformal transformations as the disformal Horndeski (DH) class.
It should be noted that one could generate ghost-free scalar-tensor theories by noninvertible conformal/disformal transformations.
However, as in generic DHOST theories that lie outside the DH class, they do not admit viable cosmology~\cite{Takahashi:2017pje,Langlois:2018jdg}.
This is another reason why we focus on invertible transformations.

Interestingly, one can include higher derivatives of the scalar field in the transformation law and still make the transformation invertible~\cite{Takahashi:2021ttd}.
In Ref.~\cite{Takahashi:2022mew}, we performed this generalized disformal transformation on Horndeski theories to construct a class of ghost-free scalar-tensor theories beyond the DHOST class, which we dubbed the generalized disformal Horndeski (GDH) class.\footnote{In general, GDH theories possess one dynamical scalar DOF on top of the dynamical metric. 
However, if the seed theory is chosen to be a special subclass of Horndeski theories where the scalar field is nondynamical (e.g., the cuscuton~\cite{Afshordi:2006ad} or its extension~\cite{Iyonaga:2018vnu,Iyonaga:2020bmm}), then the scalar field remains nondynamical after the generalized disformal transformation.}
Moreover, one could perform the generalized disformal transformation on U-DHOST theories for a further extension of ghost-free theories, at the price of introducing a shadowy mode.
We shall refer to the class of theories obtained in this way as the generalized disformal unitary-degenerate (GDU) class.
The GDH class or the GDU class provides the most general class of ghost-free scalar-tensor theories to this date.

When scalar-tensor theories are applied to cosmology, a useful and robust framework is the so-called effective field theory (EFT) of inflation/dark energy, which allows us to study cosmological perturbations in a model-independent manner~\cite{Creminelli:2006xe,Cheung:2007st,Gubitosi:2012hu}.\footnote{Note that the EFT of Refs.~\cite{Creminelli:2006xe,Cheung:2007st,Gubitosi:2012hu} assumes that a scalar field with a timelike background profile spontaneously breaks the time diffeomorphism, and hence it applies only to scalar-tensor theories. Depending on the symmetry breaking pattern, we obtain different EFTs (see, e.g., Refs.~\cite{Aoki:2021wew,Aoki:2022ipw}). Moreover, the construction of EFTs relies on the symmetry of the background spacetime. For instance, applications to anisotropic inflation~\cite{Abolhasani:2015cve,Rostami:2017wiy,Gong:2019hwj} and black hole perturbations~\cite{Franciolini:2018uyq,Hui:2021cpm,Mukohyama:2022enj,Mukohyama:2022skk,Khoury:2022zor} have been studied.}
The EFT is characterized by some time-dependent coefficients at each order of perturbative and derivative expansions, and each concrete scalar-tensor theory amounts to some particular choice of the EFT coefficients.
In particular, Horndeski theories can be described by the EFT at the leading order in the derivative expansion.
Then, the authors of Ref.~\cite{Langlois:2017mxy} proposed an extension of the EFT that accommodates DHOST theories.
The extended EFT contains several additional time-dependent coefficients associated with the higher-derivative interactions that do not appear in Horndeski theories, and their effects on the cosmic microwave background (CMB) were discussed in Refs.~\cite{Hiramatsu:2020fcd,Hiramatsu:2022fgn}.
The aim of the present paper is to further extend the EFT framework so that it incorporates GDH theories as well as GDU theories.
We also apply our EFT to study scalar and tensor cosmological perturbations.

The rest of this paper is organized as follows.
In \S\ref{sec:GDH}, we briefly review the generalized disformal transformation and define GDH/GDU theories.
In \S\ref{sec:EFT}, we construct the EFT that accommodates GDH theories.
We shall argue that the extended EFT applies also to GDU theories in Appendix~\ref{AppA} and discuss how to take into account matter fields in Appendix~\ref{AppB}.
Then, based on the EFT, we investigate scalar and tensor cosmological perturbations in \S\ref{sec:quad_ac_tensor_scalar}.
Finally, we draw our conclusions in \S\ref{sec:conc}.

\section{Generalized disformal Horndeski theories}\label{sec:GDH}

The authors of the present paper have developed a higher-derivative generalization of invertible disformal transformations in Ref.~\cite{Takahashi:2021ttd} and performed it on Horndeski theories to construct GDH theories~\cite{Takahashi:2022mew}, which form the largest class of ghost-free scalar-tensor theories known so far.
In the present paper, we focus on a subclass of GDH theories that are generated by generalized disformal transformations of the following form:
    \be
    g_\mn\quad\to\quad
    \mybar{g}_\mn[g,\phi]=f_0g_\mn+f_1\phi_\mu\phi_\nu+2f_2\phi_{(\mu}\DX{\nu)}, \qquad
    \DX{\mu}\coloneqq \bra{\delta_\mu^\alpha-\fr{\phi_\mu\phi^\alpha}{X}}\pa_\alpha X,
    \label{GDT_consistent}
    \ee
where $\phi_\mu\coloneqq \pa_\mu\phi$, $X\coloneqq \phi_\alpha\phi^\alpha$ and $f_i$'s are functions of $(\phi,X,\mZ)$, with
    \be
    \mZ\coloneqq \DX{\alpha}\DXup{\alpha}.
    \ee
As clarified in Refs.~\cite{Takahashi:2021ttd,Takahashi:2022ctx}, under a certain set of conditions, a transformation of this form has the inverse transformation at least locally in the configuration space.
We note in passing that the conventional disformal transformation~\eqref{conv_disformal} corresponds to $f_0=f_0(\phi,X)$, $f_1=f_1(\phi,X)$, and $f_2=0$.
The reason why we focus on this particular subclass is the absence of Ostrogradsky ghost in matter-coupled GDH theories:
The authors of Refs.~\cite{Takahashi:2022mew,Naruko:2022vuh,Takahashi:2022ctx} showed that both bosonic and fermionic matter fields can be consistently coupled to GDH theories.

Precisely speaking, the analyses of Refs.~\cite{Takahashi:2022mew,Naruko:2022vuh,Takahashi:2022ctx} are performed in the unitary gauge, and hence there still remains a shadowy mode in general. If one requires the absence of the shadowy mode, the generalized disformal transformation reduces to the conventional one~\cite{Ikeda:2023ntu}, restricting the GDH theories to the DH subclass.
However, the shadowy mode itself is not a propagating ghost mode, and it is harmless given an appropriate boundary condition~\cite{DeFelice:2018mkq,DeFelice:2021hps}.
Hence, the GDH theories serve as the most general framework of ghost-free scalar-tensor theories beyond the DH class.
Note that one can perform the generalized disformal transformation on U-DHOST theories~\cite{DeFelice:2018mkq,DeFelice:2021hps,DeFelice:2022xvq} to obtain a class of ghost-free scalar-tensor theories which includes the GDH class as a strict subset (see Appendix~\ref{AppA}).
Such theories, which we dub generalized disformal unitary-degenerate (GDU) theories, are also within the scope of the present paper.
To reiterate, the GDH theories satisfy the degeneracy condition in a covariant manner and the shadowy mode can show up in the presence of matter coupling.
For the GDU theories, the degeneracy condition is imposed in the unitary gauge, and hence in general there exists a shadowy mode away from the unitary gauge even in the absence of matter coupling.
The EFT framework which we shall develop in the present paper accommodates these generalized disformal theories.

Let us summarize several formulae associated with the generalized disformal metric~\eqref{GDT_consistent}.
The inverse disformal metric~$\mybar{g}^\mn$ is given by~\cite{Takahashi:2022mew,Takahashi:2022ctx}
    \be
    \mybar{g}^\mn[g,\phi]=\fr{1}{f_0}g^\mn-\fr{f_0f_1-\mZ f_2^2}{f_0\mF}\phi^\mu\phi^\nu-\fr{2f_2}{\mF}\phi^{(\mu}\DXup{\nu)}+\fr{Xf_2^2}{f_0\mF}\DXup{\mu}\DXup{\nu},
    \label{inverse_metric_spatial}
    \ee
with the quantity~$\mF$ defined by
    \be
    \mF:=f_0\bra{f_0+Xf_1}-X\mZ f_2^2. \label{calF}
    \ee
As a result, we obtain the barred counterpart of $X$ as
    \be
    \mybar{X}\coloneqq \mybar{g}^\mn\phi_\mu\phi_\nu
    =\fr{Xf_0}{\mF},
    \label{Xbar}
    \ee
which is a function of $(\phi,X,\mZ)$ in general.
In order for the disformal transformation to be invertible (i.e., $g_\mn$ can be uniquely expressed in terms of $\mybar{g}_\mn$ and $\phi$ at least locally), we require the following set of conditions~\cite{Takahashi:2022mew,Takahashi:2022ctx}:
    \be
    f_0\ne 0, \qquad
    \mF\ne 0, \qquad
    \mybar{X}_X\ne 0, \qquad
    \mybar{X}_\mZ=0, \qquad
    \bra{\fr{\mZ}{\mF}}_\mZ\ne 0, \label{inv_cond_spatial}
    \ee
where subscripts~$X$ and $\mZ$ denote the derivatives with respect to $X$ and $\mZ$, respectively.
Here, the condition~$\mybar{X}_\mZ=0$ implies that $\mybar{X}$ must be a function only of $(\phi,X)$, which puts a tight constraint among $f_i$'s through \eqref{Xbar}.
Therefore, instead of $f_i$'s, it is useful to regard $f_0(\phi,X,\mZ)$, $f_2(\phi,X,\mZ)$, and $\mybar{X}(\phi,X)$ as given functions so that the condition~$\mybar{X}_\mZ=0$ is automatically implemented.
Then, \eqref{calF} and \eqref{Xbar} allow us to express $f_1$ as
    \be
    f_1=\fr{1}{\mybar{X}}-\fr{f_0}{X}+\fr{\mZ f_2^2}{f_0}.
    \ee

In the next section, we study how the conventional EFT of inflation/dark energy~\cite{Creminelli:2006xe,Cheung:2007st,Gubitosi:2012hu} should be extended in order to accommodate GDH theories generated by the transformation~\eqref{GDT_consistent}.
Interestingly, it turns out that the extended EFT applies to GDU theories as well.

\section{Quadratic action for cosmological perturbations}\label{sec:EFT}

The EFT of inflation/dark energy~\cite{Creminelli:2006xe,Cheung:2007st,Gubitosi:2012hu} provides a model-independent framework for perturbations on a homogeneous and isotropic cosmological background.
In this context, we assume that the background scalar field~$\phi$ has a timelike profile so that one can choose the unitary gauge where $\phi=\phi(t)$ and the scalar DOF is eaten by the metric.
Note that such a scalar field spontaneously breaks the time diffeomorphism and the residual spacetime symmetries are only the spatial diffeomorphisms.
Therefore, in the unitary gauge, the action is written in terms of geometrical quantities that respect spatial covariance as well as those respecting full spacetime covariance.
In order to write down the EFT action, we introduce the Arnowitt-Deser-Misner~(ADM) variables as
    \be
    g_\mn \D x^\mu \D x^\nu=-N^2\D t^2+\ga_{ij}(\D x^i+N^i\D t)(\D x^j+N^j\D t),
    \ee
where $N$ is the lapse function, $N^i$ is the shift vector, and $\ga_{ij}$ is the spatial metric.
On the spatially flat Friedmann-Lema{\^i}tre-Robertson-Walker (FLRW) background, we have
    \be
    N=\hat{N}(t)+\delta N, \qquad
    \ga_{ij}=a(t)^2\delta_{ij}+\delta\ga_{ij},
    \ee
with $a(t)$ being the scale factor, and $N^i$ itself is a perturbed quantity.
Here and in what follows, quantities with a hat indicate their background value, though we omit the hat when unnecessary.
We denote the unit normal vector and the projection tensor associated with a constant-$t$ hypersurface respectively as
    \be
    n_\mu\coloneqq -N\delta^0_\mu, \qquad
    h_\mn\coloneqq g_\mn +n_\mu n_\nu. \label{projectiontensor}
    \ee
The extrinsic curvature and the acceleration vector are defined by
    \be
    K_{\mu\nu}\coloneqq h_\mu{}^\alpha \na_\alpha n_\nu, \qquad
    a_\mu\coloneqq n^\alpha \na_\alpha n_\mu,
    \ee
respectively, which are written in terms of the ADM variables as
    \be
    K_{ij}=\fr{1}{2N}\bra{\dot{\ga}_{ij}-{\rm D}_iN_j-{\rm D}_jN_i}, \qquad
    a_i=\fr{1}{N}{\rm D}_iN,
    \label{Kij,ai}
    \ee
with a dot denoting the derivative with respect to $t$ and ${\rm D}_i$ being the covariant derivative associated with $\ga_{ij}$.
The spatial curvature tensor~$\Rs_{\mu\nu\lambda\sigma}$ associated with $\ga_{ij}$ can also be a building block of the EFT action.
Note that the spatial curvature tensor can be written as a combination of the four-dimensional spacetime curvature tensor and the extrinsic curvature through the Gauss relation, and hence one can employ either the spatial or spacetime curvature as an independent building block.
The EFT action is then written in terms of these objects, i.e.,
    \be
    S=\int \D^4x\sqrt{-g}\,\mL(R_{\mu\nu\lambda\sigma},g^{00},K_\mn,\nabla_\mu,t).
    \ee
Practically, one performs perturbative and derivative expansions up to the necessary order.

Given an action of scalar-tensor theories, it is straightforward to recast it in the form of the EFT action.
Indeed, under the unitary gauge, the first and second derivatives of $\phi$ are related to the above geometrical quantities as~\cite{Gleyzes:2013ooa}
    \be
    \phi_\mu=-\fr{\dot{\phi}}{N}n_\mu, \qquad
    \phi_{\mu\nu}=-\fr{\dot{\phi}}{N}(K_{\mu\nu}-n_\mu a_\nu-n_\nu a_\mu)-\fr{N}{2\dot{\phi}}(n^\alpha \pa_\alpha X)n_\mu n_\nu,
    \label{phi_der_unitary}
    \ee
with $\phi_\mn\coloneqq \na_\mu\na_\nu\phi$.
From the first equation, we see that $X=\phi_\alpha\phi^\alpha=-\dot{\phi}^2/N^2$ and
    \be
    \begin{split}
    \DX{\mu}&=-2Xa_\mu=-\fr{2X}{N}\delta_\mu^i\pa_i\delta N+\mO(\delta^2), \\
    \mZ&=4X^2a_\mu a^\mu=4X^2\fr{(\pa_i\delta N)^2}{N^2a^2}+\mO(\delta^3),
    \end{split} \label{calX_unitary}
    \ee
where $\mO(\delta^n)$ denotes terms of $n$th or higher order in perturbations.
Note that both $\DX{\mu}$ and $\mZ$ vanish at the background level.

In \S\ref{ssec:seed_action}, we first review how the Horndeski class is embedded in the EFT framework.
The EFT action derived here is regarded as the seed of the generalized disformal transformation.
Then, we study how each building block of the seed EFT action is transformed under the generalized disformal transformation~\eqref{GDT_consistent} in \S\ref{ssec:building_block}.
Finally, combining the results of \S\ref{ssec:seed_action} and \S\ref{ssec:building_block}, we obtain the EFT action that accommodates GDH theories in \S\ref{ssec:GDH_EFT}.
For demonstration purposes, we focus on the action up to the quadratic order in perturbations in the present paper, but our discussion can be applied to cubic or any higher-order actions in principle.

\subsection{Seed action}\label{ssec:seed_action}

The Horndeski action up to the quadratic order in the second derivative of the scalar field is given by
    \be
    S_{\rm H}=\int \D^4x\sqrt{-g}\,\mL, \qquad
    \mL\coloneqq G_2+G_3\Box\phi+G_4R-2G_{4X}\brb{(\Box\phi)^2-\phi^\mu_\nu\phi_\mu^\nu}, \label{S_H}
    \ee
with $G_I$'s being functions of $(\phi,X)$.\footnote{For simplicity, we do not include the so-called quintic Horndeski Lagrangian.}
In the unitary gauge, one can employ \eqref{phi_der_unitary} to recast the Lagrangian in the form
    \be
    \mL(t,N,K,K_2,\Rs)=b_0+b_1K+b_2K_2+b_3\Rs,
    \label{L_H_unitary}
    \ee
where $K\coloneqq K^i_i$, $K_2\coloneqq K^2-K^i_jK_i^j$, and $\Rs$ denotes the spatial Ricci scalar.
Here, $b_0$, $b_1$, $b_2$, and $b_3$ are functions of $(t,N)$, which are related to $G_I$'s through~\cite{Gleyzes:2013ooa}
    \be
    b_0=G_2-Xg_{3\phi}, \qquad
    b_1=-2\sqrt{-X}\bra{Xg_{3X}+G_{4\phi}}, \qquad
    b_2=-G_4+2XG_{4X}, \qquad
    b_3=G_4.
    \ee
Here, the function~$g_3(\phi,X)$ has been defined so that $g_3+2Xg_{3X}=G_3$.

Let us study perturbations about the spatially flat FLRW background spacetime.
We define the perturbation of the extrinsic curvature as
    \be
    \delta K^i_j\coloneqq K^i_j-H\delta^i_j, \qquad
    \delta K\coloneqq K-3H,
    \ee
where $H\coloneqq \dot{a}/(\hat{N}a)$ is the Hubble parameter.
Note that
    \be
    \delta K_2\coloneqq K_2-6H^2
    =4H\delta K+\delta K^2-\delta K^i_j\delta K_i^j+\mO(\delta^3).
    \ee
Note also that $\Rs$ vanishes at the background level for the spatially flat FLRW metric, and we write $\delta\Rs=\Rs$.
Then, one can expand the Lagrangian~\eqref{L_H_unitary} up to the quadratic order in perturbations to rewrite the Horndeski action as
    \begin{align}
    S_{\rm H}
    =\int \D^4x\sqrt{-g}
    \Bigg[&\hat{\mL}-3H{\cal A}-\fr{\dot{\cal A}}{\hat{N}}+\bra{\hat{N}\hat{\mL}_{N}+\fr{\dot{\cal A}}{\hat{N}}}\fr{\delta N}{\hat{N}}+\hat{\mL}_{\Rs}\delta\Rs-\hat{\mL}_{K_2}\bra{\delta K^i_j\delta K^j_i-\delta K^2} \nonumber \\
    &+\bra{\fr{\hat{N}^2}{2}\hat{\mL}_{NN}-\fr{\dot{\cal A}}{\hat{N}}}\fr{\delta N^2}{\hat{N}^2}+\bra{\hat{\mL}_{NK}+4H\hat{\mL}_{NK_2}}\delta N\delta K+\hat{\mL}_{N\Rs}\delta N\delta\Rs+{\cal O} (\delta^3)\Bigg], \label{S_H_unitary}
    \end{align}
where we have defined
    \be
    {\cal A}(t)\coloneqq \hat{\mL}_K+4H\hat{\mL}_{K_2}.
    \ee
Here, $\hat{\mL}$ with subscript(s) denotes the derivative of the Lagrangian~\eqref{L_H_unitary} with respect to the subscript(s) evaluated at the background.
One can employ the background equations of motion (EOMs) to simplify the above action.
In the absence of matter fields, the background EOMs read
    \be
    \mE_0\coloneqq \hat{\mL}-3H{\cal A}-\fr{\dot{\cal A}}{\hat{N}}=0, \qquad
    \mE_1\coloneqq \hat{N}\hat{\mL}_{N}+\fr{\dot{\cal A}}{\hat{N}}=0,
    \label{BGEOMs}
    \ee
which can be used to remove the following part from the action~\eqref{S_H_unitary}:
    \be
    S_0
    =\int \D^4x\sqrt{-g}\brb{\mE_0+\mE_1\bra{\fr{\delta N}{\hat{N}}-\fr{\delta N^2}{\hat{N}^2}}}. \label{S_0}
    \ee
After using the background EOMs, the quadratic action in a cosmological background can be written as
    \begin{align}
    \delta_2S_{\rm H}
    =\int \D t\D^3x\,\hat{N}a^3
    \bigg[&-\hat{\mL}_{K_2}\bra{\delta_1 K^i_j\delta_1 K^j_i-\delta_1 K^2}+\hat{\mL}_{\Rs}\delta_2\bra{\Rs\fr{\sqrt{\gamma}}{a^3}}+\bra{\fr{1}{2}\hat{\mL}_{NN}+\fr{\hat{\mL}_N}{\hat{N}}}\delta_1 N^2 \nonumber \\
    &+\bra{\hat{\mL}_{NK}+4H\hat{\mL}_{NK_2}}\delta_1 N\delta_1 K+\bra{\hat{N}\hat{\mL}_{N\Rs}+\hat{\mL}_{\Rs}}\fr{\delta_1 N}{\hat{N}}\delta_1\!\Rs\bigg].
    \label{S_H_quad}
    \end{align}
Here and in what follows, $\delta_1 Q$ and $\delta_2 Q$ denote the parts of a quantity~$Q$ that are of linear and quadratic orders in perturbations, respectively.

For phenomenological purposes, the following parametrization is often used to describe the quadratic action up to the leading order in the derivative expansion~\cite{Bellini:2014fua,Gleyzes:2014qga,Gleyzes:2014rba,Frusciante:2016xoj}:
    \begin{align}
    \delta_2S_{\rm g}
    =\int \D t\D^3x\,\hat{N}a^3\fr{\ti{M}^2}{2}
    \bigg[&\delta_1 K^i_j\delta_1 K^j_i-\bra{1+\fr{2}{3}\aH_{\rm L}}\delta_1 K^2+(1+\aH_{\rm T})\delta_2\bra{\Rs\fr{\sqrt{\gamma}}{a^3}} \nonumber \\
    &+H^2\aH_{\rm K}\fr{\delta_1 N^2}{\hat{N}^2}+4H\aH_{\rm B}\fr{\delta_1 N}{\hat{N}}\delta_1 K+(1+\aH_{\rm H})\fr{\delta_1 N}{\hat{N}}\delta_1\!\Rs\bigg],
    \label{EFT_lowest}
    \end{align}
where $\ti{M}$, $\aH_{\rm L}$, $\aH_{\rm T}$, $\aH_{\rm K}$, $\aH_{\rm B}$, and $\aH_{\rm H}$ are functions of $t$.
Comparing this equation with \eqref{S_H_quad}, we see that the Horndeski Lagrangian~\eqref{L_H_unitary} amounts to the following choice of parameters:
    \be
    \begin{split}
    &\ti{M}^2=-2\hat{\mL}_{K_2}, \qquad
    \aH_{\rm L}=0, \qquad
    \aH_{\rm T}=\fr{2\hat{\mL}_{\Rs}}{M^2}-1, \qquad
    \aH_{\rm K}=\fr{\hat{N}}{M^2H^2}\bra{\hat{N}\hat{\mL}_{NN}+2\hat{\mL}_N}, \\
    &\aH_{\rm B}=\fr{\hat{N}}{2M^2H}\bra{\hat{\mL}_{NK}+4H\hat{\mL}_{NK_2}}, \qquad
    \aH_{\rm H}=0.
    \end{split}
    \ee
Equivalently, in terms of the coefficient functions~$G_I$ in the original action~\eqref{S_H}, we have
    \be
    \begin{split}
    &\ti{M}^2=\mG_T, \qquad
    \aH_{\rm L}=0, \qquad
    \aH_{\rm T}=\fr{4XG_{4X}}{\mG_T}, \qquad
    \aH_{\rm K}=\fr{2\Sigma+12H\Theta-6H^2\mG_T}{H^2\mG_T}, \\
    &\aH_{\rm B}=\fr{\Theta-H\mG_T}{H\mG_T}, \qquad
    \aH_{\rm H}=0.
    \end{split} \label{alpha_param_Horndeski}
    \ee
Here, we have defined
    \be
    \begin{split}
    \mG_T&\coloneqq 2(G_4-2XG_{4X}), \\
    \Sigma&\coloneqq XG_{2X}+2X^2G_{2XX}
    -6H\dot{\phi}\bra{2XG_{3X}+X^2G_{3XX}}-X\bra{G_{3\phi}+XG_{3\phi X}} \\
    &\quad -6H^2\bra{G_4-7XG_{4X}-16X^2G_{4XX}-4X^3G_{4XXX}}-6H\dot{\phi}\bra{G_{4\phi}+5XG_{4\phi X}+2X^2G_{4\phi XX}}, \\
    \Theta&\coloneqq \dot{\phi}XG_{3X}+2H\bra{G_{4}-4XG_{4X}-4X^2G_{4XX}}+\dot{\phi}\bra{G_{4\phi}+2XG_{4\phi X}},
    \end{split} \label{SigmaTheta}
    \ee
where the right-hand sides are evaluated at the background.

It should be noted that the quadratic action~\eqref{EFT_lowest} also applies to U-DHOST theories.
Indeed, there exists a canonical form of quadratic U-DHOST theories, for which the quadratic action for cosmological perturbations boils down to the form~\eqref{EFT_lowest} (see Appendix~\ref{AppA}).
In contrast to the case of Horndeski theories, $\aH_{\rm L}$ and $\aH_{\rm H}$ can be nonvanishing in U-DHOST theories.
Note also that, in rewriting the quadratic action in the form~\eqref{EFT_lowest}, we removed the terms in \eqref{S_0} by use of the vacuum background EOMs~\eqref{BGEOMs}.
In the presence of matter fields, this procedure should be modified accordingly.
We provide a prescription for how to incorporate the contribution from matter fields in Appendix~\ref{AppB}.

\subsection{Transformation of the EFT building blocks}\label{ssec:building_block}

In this subsection, we study the transformation law for each building block of the leading-order EFT action~\eqref{EFT_lowest} under the generalized disformal transformation~\eqref{GDT_consistent}.
As in \eqref{GDT_consistent}, quantities after the transformation are indicated with a bar.
Since we focus on the quadratic action, we omit terms of higher order in perturbations.

\begin{itemize}
\item {\it ADM variables}.
The transformation law for the ADM variables is given as follows \cite{Takahashi:2022mew}:
    \be
    \mybar{N}^2=-\fr{\dot{\phi}^2}{\mybar{X}(\phi,-\dot{\phi}^2/N^2)}, \qquad
    \mybar{N}_i=f_0N_i+\dot{\phi}f_2\DX{i}, \qquad
    \mybar{\gamma}_{ij}=f_0\gamma_{ij}.
    \label{barred_ADM}
    \ee
In particular, $\mybar{\gamma}_{ij}$ is nothing but a conformal transformation of $\gamma_{ij}$.
Note that we have assumed that the metric signature is preserved under the generalized disformal transformation, which is realized if $\mybar{X}/X>0$.
From \eqref{barred_ADM}, we find that the background value of the lapse function and the scale factor are transformed as
    \be
    \hat{\mybar{N}}=\bra{\fr{\hat{X}}{\hat{\mybar{X}}}}^{1/2}\hat{N}, \qquad
    \mybar{a}=\hat{f}_0^{1/2}a.
    \ee
Moreover, we have
    \be
    \begin{split}
    \fr{\delta\mybar{N}}{\mybar{N}}&=\fr{X\mybar{X}_X}{\mybar{X}}\fr{\delta N}{N}+\mO(\delta^2), \\
    \mybar{N}_i&=f_0N_i-2\dot{\phi}Xf_2\fr{\pa_i\delta N}{N}+\mO(\delta^3).
    \end{split}
    \ee
The following relation is also useful:
    \be
    \fr{\delta\sqrt{\mybar{\gamma}}}{\mybar{a}^3}
    =\fr{\delta\sqrt{\gamma}}{a^3}-\fr{3Xf_{0X}}{f_0}\fr{\delta N}{N}+\mO(\delta^2).
    \ee

\item {\it Extrinsic curvature}.
The barred counterpart of the extrinsic curvature is defined by
    \begin{align}
    \mybar{K}_{ij}=\fr{1}{2\mybar{N}}\bra{\dot{\mybar{\ga}}_{ij}-\mybar{\rm D}_i\mybar{N}_j-\mybar{\rm D}_j\mybar{N}_i},
    \end{align}
and hence
    \begin{align}
    \mybar{K}^i_j\coloneqq \mybar{\gamma}^{ik}\mybar{K}_{kj}
    =\fr{1}{f_0\mybar{N}}\bra{N f_0K^i_j+\fr{1}{2}\dot{f}_0\delta^i_j-\dot{\phi}f_2{\rm D}^i\DX{j}}+\mO(\delta^2).
    \end{align}
Therefore, the perturbation of the extrinsic curvature is transformed as
    \be
    \begin{split}
    \delta \mybar{K}^i_j&=\bra{\fr{\mybar{X}}{X}}^{1/2}\brb{\delta K^i_j+H\mB\fr{\delta N}{N}\delta^i_j+\fr{\mC}{N}\bra{\fr{\delta N}{N}}^{\boldsymbol{\cdot}}\delta^i_j+\fr{\mD}{\MD}\fr{\pa^i\pa_j\delta N}{Na^2}}+\mO(\delta^2), \\
    \delta \mybar{K}&=\bra{\fr{\mybar{X}}{X}}^{1/2}\brb{\delta K+3H\mB\fr{\delta N}{N}+\fr{3\mC}{N}\bra{\fr{\delta N}{N}}^{\boldsymbol{\cdot}}+\fr{\mD}{\MD}\fr{\Delta\delta N}{Na^2}}+\mO(\delta^2),
    \end{split}
    \ee
where $\pa^i=\pa_i$, $\Delta\coloneqq \pa_i^2$, and we have defined the following dimensionless quantities:
    \be
    \mB\coloneqq \left.\fr{1}{NH}\brb{\sqrt{-X}\bra{\sqrt{-X}\ln f_0}_X}^{\boldsymbol{\cdot}}+\fr{2\mybar{H}}{H}X\pa_X\sqrt{\fr{X}{\mybar{X}}}\,\right|_{\rm BG}, \qquad
    \mC\coloneqq -\fr{\hat{X}\hat{f}_{0X}}{\hat{f}_0}, \qquad
    \mD\coloneqq -\fr{2\MD(-\hat{X})^{3/2}\hat{f}_2}{\hat{f}_0}.
    \label{BCD}
    \ee
Here, $\mybar{H}\coloneqq \dot{\mybar{a}}/\hat{\mybar{N}}\mybar{a}$ and $\MD$ is some constant mass scale that characterizes the energy scale associated with $f_2$.
It should be noted that there is a priori no typical scale for $\MD$.
This is because the disformal factor~$f_2$ multiplies to the term~$\phi_{(\mu}\DX{\nu)}$ in the generalized disformal transformation~\eqref{GDT_consistent}, which does not have a background value on a homogeneous and isotropic background [see Eq.~\eqref{calX_unitary}].

\item {\it Spatial curvature}.
Since the spatial metric is transformed conformally, the spatial curvature is transformed as
    \be
    \bRs=\fr{1}{f_0}\Rs
    +\fr{3}{2f_0^3}{\rm D}_kf_0{\rm D}^kf_0
    -\fr{2}{f_0^2}{\rm D}_k{\rm D}^kf_0,
    \ee
which holds exactly.
The following relations are useful:
    \be
    \begin{split}
    \delta_1\!\bRs&=\fr{1}{\hat{f}_0}\delta_1\Rs-\fr{4\mC}{\hat{f}_0}\fr{\Delta\delta_1 N}{\hat{N}a^2}, \\
    \delta_2\bra{\bRs\fr{\sqrt{\mybar{\gamma}}}{\mybar{a}^3}}
    &=\fr{1}{\hat{f}_0}\delta_2\bra{\Rs\fr{\sqrt{\gamma}}{a^3}}+\fr{\mC}{\hat{f}_0}\fr{\delta_1 N}{\hat{N}}\delta_1\!\Rs+\fr{2\mC^2}{\hat{f}_0}\fr{(\pa_i\delta_1 N)^2}{\hat{N}^2a^2}+\bra{\text{spatial total derivatives}},
    \end{split}
    \ee
where in the second equation we have omitted spatial total derivatives of $\mO(\delta^2)$ as they do not contribute to the quadratic action.
\end{itemize}

\subsection{Quadratic action for generalized disformal Horndeski theories}\label{ssec:GDH_EFT}

Having developed the transformation law for each EFT building block, we can now discuss the transformation of the seed quadratic action~\eqref{EFT_lowest}.
After straightforward manipulations, we arrive at the following expression:
    \begin{align}
    \delta_2S_{\rm g}
    =\int \D t\D^3x\,\hat{N}a^3\fr{M^2}{2}
    \bigg\{&\delta_1 K^i_j\delta_1 K^j_i-\bra{1+\fr{2}{3}\alpha_{\rm L}}
    \delta_1 K^2+(1+\alpha_{\rm T})\delta_2\bra{\Rs\fr{\sqrt{\gamma}}{a^3}}
    +H^2\alpha_{\rm K}\fr{\delta_1 N^2}{\hat{N}^2}+4H\alpha_{\rm B}\fr{\delta_1 N}{\hat{N}}\delta_1 K \nonumber \\
    &+(1+\alpha_{\rm H})\fr{\delta_1 N}{\hat{N}}\delta_1\!\Rs
    +\fr{4\beta_1}{\hat{N}}\bra{\fr{\delta_1 N}{\hat{N}}}^{\boldsymbol{\cdot}}\delta_1 K+\fr{\beta_2}{\hat{N}^2}\brb{\bra{\fr{\delta_1 N}{\hat{N}}}^{\boldsymbol{\cdot}}\,}^2+\beta_3\fr{(\pa_i\delta_1 N)^2}{\hat{N}^2a^2} \nonumber \\
    &+\fr{\gamma_1}{\MD}\delta_1 K^i_j\fr{\pa_i\pa^j\delta_1 N}{\hat{N}a^2}+\fr{\gamma_2}{\MD}\delta_1 K\fr{\Delta\delta_1 N}{\hat{N}a^2}+\fr{\gamma_3}{\MD^2}\fr{(\Delta\delta_1 N)^2}{\hat{N}^2a^4}\bigg\},
    \label{EFT_GDT}
    \end{align}
where $M$, $\alpha$'s, $\beta$'s, and $\gamma$'s are functions of $t$ given by
    \be
    \begin{split}
    &M^2=\bra{\fr{\mybar{X}f_0^3}{X}}^{1/2}\ti{M}^2, \qquad
    \alpha_{\rm L}=\aH_{\rm L}, \qquad
    \alpha_{\rm T}=\fr{X}{\mybar{X}f_0}(1+\aH_{\rm T})-1, \\
    &\alpha_{\rm H}=\fr{X^2\mybar{X}_X}{\mybar{X}^2f_0}(1+\aH_{\rm H})+\fr{X\mC}{\mybar{X}f_0}(1+\aH_{\rm T})-1, \qquad
    \alpha_{\rm B}=\fr{\mybar{H}}{H}\bra{\fr{X}{\mybar{X}}}^{3/2}\mybar{X}_X\aH_{\rm B}-\mB(1+\aH_{\rm L}), \\
    &\alpha_{\rm K}=\fr{\mybar{H}^2}{H^2}\fr{X^3\mybar{X}_X^2}{\mybar{X}^3}\aH_{\rm K}+6\mB\brb{\mB(1+\aH_{\rm L})+2\alpha_{\rm B}}
    -\fr{6(a^3H\mC M^2\alpha_{\rm B})^{\boldsymbol{\cdot}}}{Na^3H^2M^2}, \\
    &\beta_1=-\mC(1+\aH_{\rm L}), \qquad
    \beta_2=-6\mC^2(1+\aH_{\rm L}), \\
    &\beta_3=\fr{4X^2\mybar{X}_X\mC}{\mybar{X}^2f_0}(1+\aH_{\rm H})+\fr{2X\mC^2}{\mybar{X}f_0}(1+\aH_{\rm T})
   -\fr{4H\mD\alpha_{\rm B}}{\MD}+\fr{2(a\mD M^2\beta_1)^{\boldsymbol{\cdot}}}{NaM^2\MD}, \\
    &\gamma_1=2\mD, \qquad
    \gamma_2=-2\mD\bra{1+\fr{2}{3}\aH_{\rm L}}, \qquad
    \gamma_3=-\fr{2}{3}\mD^2\aH_{\rm L},
    \end{split} \label{alpha_param_disformal}
    \ee
with $\ti{M}$ and $\aH$'s being the EFT parameters in the seed action~\eqref{EFT_lowest}.
Here, all the quantities on the right-hand sides should be evaluated at the background.
Note that the quadratic action~\eqref{EFT_GDT} with the coefficients given by \eqref{alpha_param_disformal} does not yield Ostrogradsky ghosts, since it is generated by the invertible generalized disformal transformation from the action~\eqref{EFT_lowest} which does not contain higher time derivatives.
The action~\eqref{EFT_GDT} itself makes sense even if it were not disformally related to \eqref{EFT_lowest}, but in this case there would be an Ostrogradsky mode in general.

In particular, the EFT parameters for GDH theories can be obtained by substituting \eqref{alpha_param_Horndeski} into \eqref{alpha_param_disformal}.
In doing so, $X$ and $H$ in \eqref{alpha_param_Horndeski} should be replaced by $\mybar{X}$ and $\mybar{H}$, respectively.
Written explicitly,
    \be
    \begin{split}
    &M^2=\bra{\fr{\mybar{X}f_0^3}{X}}^{1/2}\mybar{\mG}_T, \qquad
    \alpha_{\rm L}=0, \qquad
    \alpha_{\rm T}=\fr{X}{\mybar{X}f_0}\fr{2G_4}{\mybar{\mG}_T}-1, \\
    &\alpha_{\rm H}=\fr{X^2\mybar{X}_X}{\mybar{X}^2f_0}+\fr{X\mC}{\mybar{X}f_0}\fr{2G_4}{\mybar{\mG}_T}-1, \qquad
    \alpha_{\rm B}=\bra{\fr{X}{\mybar{X}}}^{3/2}\mybar{X}_X\fr{\mybar{\Theta}-\mybar{H}\mybar{\mG}_T}{H\mybar{\mG}_T}-\mB, \\
    &\alpha_{\rm K}=\fr{X^3\mybar{X}_X^2}{\mybar{X}^3}\fr{2\mybar{\Sigma}+12\mybar{H}\mybar{\Theta}-6\mybar{H}^2\mybar{\mG}_T}{H^2\mybar{\mG}_T}+6\mB\bra{\mB+2\alpha_{\rm B}}
    -\fr{6(a^3H\mC M^2\alpha_{\rm B})^{\boldsymbol{\cdot}}}{Na^3H^2M^2}, \\
    &\beta_1=-\mC, \qquad
    \beta_2=-6\mC^2, \qquad
    \beta_3=\fr{4X^2\mybar{X}_X\mC}{\mybar{X}^2f_0}+\fr{2X\mC^2}{\mybar{X}f_0}\fr{2G_4}{\mybar{\mG}_T}
    -\fr{4H\mD\alpha_{\rm B}}{\MD}-\fr{2(a\mC\mD M^2)^{\boldsymbol{\cdot}}}{NaM^2\MD}, \\
    &\gamma_1=-\gamma_2=2\mD, \qquad
    \gamma_3=0,
    \end{split} \label{alpha_param_GDH}
    \ee
where
    \be
    \begin{split}
    \mybar{\mG}_T&\coloneqq 2(G_4-2\mybar{X}G_{4X}), \\
    \mybar{\Sigma}&\coloneqq \mybar{X}G_{2X}+2\mybar{X}^2G_{2XX}
    -6\mybar{H}\dot{\phi}\bra{2\mybar{X}G_{3X}+\mybar{X}^2G_{3XX}}-\mybar{X}\bra{G_{3\phi}+\mybar{X}G_{3\phi X}} \\
    &\quad -6\mybar{H}^2\bra{G_4-7\mybar{X}G_{4X}-16\mybar{X}^2G_{4XX}-4\mybar{X}^3G_{4XXX}}-6\mybar{H}\dot{\phi}\bra{G_{4\phi}+5\mybar{X}G_{4\phi X}+2\mybar{X}^2G_{4\phi XX}}, \\
    \mybar{\Theta}&\coloneqq \dot{\phi}\mybar{X}G_{3X}+2\mybar{H}\bra{G_{4}-4\mybar{X}G_{4X}-4\mybar{X}^2G_{4XX}}+\dot{\phi}\bra{G_{4\phi}+2\mybar{X}G_{4\phi X}},
    \end{split} \label{SigmaTheta_bar}
    \ee
with $G_I$'s evaluated at $(\phi,\mybar{X})$.

As mentioned previously, the quadratic action~\eqref{EFT_GDT} also applies to GDU theories (see Appendix~\ref{AppA} for details).
In this case, the parameter~$\alpha_{\rm L}$ can be nonvanishing, which is in contrast to the case of GDH theories.
Indeed, as we shall see in \S\ref{ssec:scalar_pert}, the parameter~$\alpha_{\rm L}$ controls the so-called scordatura effect~\cite{Motohashi:2019ymr} which is built-in in generic U-DHOST theories~\cite{DeFelice:2022xvq}.

Compared to the seed action~\eqref{EFT_lowest}, the action~\eqref{EFT_GDT} involves new parameters, i.e., $\beta$'s and $\gamma$'s.
Among them, $\beta_1$, $\beta_2$, and $\beta_3$ are those already present in the EFT for DHOST theories~\cite{Langlois:2017mxy}.
On the other hand, the parameters~$\gamma_1$, $\gamma_2$, and $\gamma_3$ are peculiar to our extended EFT for GDH theories.
These parameters are proportional to $\mD$ (or its squared) as they are generated by the generalized disformal transformation with nonvanishing $f_2$ [see Eq.~\eqref{BCD}].
Note also that, if we choose
    \be
    f_0=f_0(\phi), \qquad
    f_1=f_1(\phi), \qquad
    f_2=0, \qquad
    \mybar{X}=\fr{X}{f_0+Xf_1},
    \ee
the resultant theory lies within the Horndeski class.
Therefore, in this particular case, $\alpha_{\rm L}$ and $\alpha_{\rm H}$ as well as all the new parameters~$\beta$'s and $\gamma$'s vanish as in the seed Horndeski theory [see Eq.~\eqref{alpha_param_Horndeski}], as it should.

If we focus on a subclass of GDH theories with $f_2=0$, we have $\mD=0$, and hence
    \be
    \beta_2=-6\beta_1^2, \qquad
    \beta_3=-2\beta_1\brb{2(1+\alpha_{\rm H})+\beta_1(1+\alpha_{\rm T})}, \qquad
    \gamma_1=\gamma_2=\gamma_3=0.
    \ee
Interestingly, the same set of relations is satisfied 
by DH theories, which are generated from Horndeski theories by the conventional disformal transformation~\eqref{conv_disformal}.
This implies that this subclass of GDH theories cannot be distinguished from the DH class at least at the level of linear cosmological perturbations.
In other words, deviations from the DH class would show up only when $f_2\ne 0$.

\section{Quadratic actions for tensor and scalar perturbations}\label{sec:quad_ac_tensor_scalar}

In this section, we study tensor and scalar perturbations on a homogeneous and isotropic cosmological background based on the quadratic action~\eqref{EFT_GDT} of the extended EFT.
Note that vector perturbations are nondynamical in GDH theories unless the matter sector introduces a dynamical vector mode.

\subsection{Tensor perturbations}\label{ssec:tensor_pert}

Let us consider tensor perturbations given by
    \be
    \delta N=0, \qquad
    N^i=0, \qquad
    \delta\gamma_{ij}=a^2h_{ij},
    \ee
with $h_{ii}=0=\pa_j h_{ij}$.
From the EFT action~\eqref{EFT_GDT}, the quadratic action for tensor perturbations can be derived as
    \be
    \delta_2S_T
    =\int \D t\D^3x\,\hat{N}a^3\fr{M^2}{8}
    \brb{\fr{\dot{h}_{ij}^2}{\hat{N}^2}-(1+\alpha_{\rm T})\fr{(\pa_kh_{ij})^2}{a^2}}.
    \label{qac_tensor}
    \ee
Therefore, we find the squared sound speed for the tensor perturbations as
    \be
    c_T^2=1+\alpha_{\rm T}
    =\fr{X}{\mybar{X}f_0}\fr{2G_4}{\mybar{\mG}_T}.
    \label{cT}
    \ee
Note that matter fields are assumed to be minimally coupled, and hence light propagates at unit speed.
Also, the conditions for the absence of ghost/gradient instabilities are respectively given by
    \be
    M^2>0, \qquad
    1+\alpha_{\rm T}>0. \label{stability_tensor}
    \ee
From \eqref{cT}, one can read off the condition under which gravitational waves (GWs) propagate at the speed of light.
This condition makes the EFT consistent with the almost simultaneous detection of the GW event~GW170817 and the gamma-ray burst~170817A emitted from a binary neutron star merger~\cite{TheLIGOScientific:2017qsa,GBM:2017lvd,Monitor:2017mdv}.
Solving $c_T^2=1$ for $f_0$, we have
    \be
    f_0=\fr{X}{\mybar{X}}\fr{2G_4}{\mybar{\mG}_T}
    =\fr{X}{\mybar{X}}\fr{G_4(\phi,\mybar{X})}{G_4(\phi,\mybar{X})-2\mybar{X}G_{4X}(\phi,\mybar{X})}.
    \ee
This result is consistent with the condition obtained in \cite{Takahashi:2022mew}.

\subsection{Scalar perturbations}\label{ssec:scalar_pert}

Let us consider scalar perturbations given by
    \be
    \delta N=\hat{N}\dn, \qquad
    N_i=\hat{N}\pa_i\psi, \qquad
    \delta\gamma_{ij}=2a^2\bra{\zeta\delta_{ij}+\pa_i\pa_jE}.
    \ee
In what follows, we adopt the gauge $E=0$ on top of $\delta\phi=0$ which we have already imposed in constructing the EFT action~\eqref{EFT_GDT}.
Note that this is a complete gauge fixing and hence can be imposed at the action level~\cite{Motohashi:2016prk}.
We first consider only the gravitational action and the effect of matter fields will be discussed later in this subsection.
Then, the quadratic action for scalar perturbations can be written as
    \begin{align}
    \delta_2S_S
    =\int \D t\D^3x\,\hat{N}a^3\fr{M^2}{2}
    \Biggl[\!&-6(1+\alpha_{\rm L})\bra{\fr{\dot{\zeta}}{\hat{N}}-H\dn-\fr{\Delta\psi}{3a^2}}^2+\fr{2(\Delta\psi)^2}{3a^4}+2(1+\alpha_{\rm T})\fr{(\pa_i\zeta)^2}{a^2}+H^2\alpha_{\rm K}\dn^2 \nonumber \\
    &+\bra{12H\alpha_{\rm B}\dn+12\beta_1\fr{\dot{\dn}}{\hat{N}}+\fr{\gamma_1+3\gamma_2}{\MD}\fr{\Delta\dn}{a^2}}\bra{\fr{\dot{\zeta}}{\hat{N}}-H\dn-\fr{\Delta\psi}{3a^2}}-4(1+\alpha_{\rm H})\dn\fr{\Delta\zeta}{a^2} \nonumber \\
    &+\beta_2\bra{\fr{\dot{\dn}}{\hat{N}}}^2+\beta_3\fr{(\pa_i\dn)^2}{a^2}-\fr{2\gamma_1}{3\MD}\fr{\Delta\dn\Delta\psi}{a^4}+\fr{\gamma_3}{\MD^2}\bra{\fr{\Delta\dn}{a^2}}^2\,\Biggr].
    \label{qac_scalar}
    \end{align}
In general, this action involves two dynamical DOFs, one of which is the Ostrogradsky mode.
However, if the EFT~\eqref{EFT_GDT} is generated from the leading-order one~\eqref{EFT_lowest} by the generalized disformal transformation, then the Ostrogradsky mode does not show up.
In such a case, one has the following relations among the EFT parameters [see Eq.~\eqref{alpha_param_disformal}]:
    \be
    \begin{split}
    &\beta_2=-\fr{6\beta_1^2}{1+\alpha_{\rm L}}, \qquad
    \beta_3=-4\beta_1\fr{1+\alpha_{\rm H}}{1+\alpha_{\rm L}}-2\beta_1^2\fr{1+\alpha_{\rm T}}{(1+\alpha_{\rm L})^2}
    -\fr{2H\alpha_{\rm B}\gamma_1}{\MD}+\fr{(aM^2\beta_1\gamma_1)^{\boldsymbol{\cdot}}}{\hat{N}aM^2\MD}, \\
    &\gamma_2=-\bra{1+\fr{2}{3}\alpha_{\rm L}}\gamma_1, \qquad
    \gamma_3=-\fr{\alpha_{\rm L}}{6}\gamma_1^2
    =\fr{1}{4}\gamma_1(\gamma_1+\gamma_2).
    \end{split} \label{EFT-param_relation}
    \ee
In particular, the first relation ensures the absence of Ostrogradsky mode, and hence there is only one dynamical DOF.
After redefining $\zeta$ as
    \be
    \ti{\zeta}\coloneqq \zeta-\fr{\beta_1}{1+\alpha_{\rm L}}\dn,
    \label{tilde_zeta}
    \ee
and imposing \eqref{EFT-param_relation}, the action takes the form
    \begin{align}
    \delta_2S_S
    =\int \D t\D^3x\,\hat{N}a^3\fr{M^2}{2}
    \Biggl[&c_0\bra{\fr{\dot{\ti{\zeta}}}{\hat{N}}}^2+\fr{\dot{\ti{\zeta}}}{\hat{N}}\bra{c_1-c_2\fr{\Delta}{a^2}}\dn-c_3\fr{\dot{\ti{\zeta}}}{\hat{N}}\fr{\Delta\psi}{a^2}+c_4\fr{(\pa_i\ti{\zeta})^2}{a^2}-c_5\dn\fr{\Delta\ti{\zeta}}{a^2} \nonumber \\
    &+\dn\bra{c_6-c_7\fr{\Delta}{a^2}+c_8\fr{\Delta^2}{a^4}}\dn-\nu\bra{c_9-c_{10}\fr{\Delta}{a^2}}\fr{\Delta\psi}{a^2}+c_{11}\bra{\fr{\Delta\psi}{a^2}}^2\,\Biggr],
    \label{qac_scalar2}
    \end{align}
where the coefficients are given by
    \be
    \begin{split}
    &c_0=-6(1+\alpha_{\rm L}), \qquad
    c_1=12H\brb{1+\alpha_{\rm B}+\alpha_{\rm L}-\fr{1+\alpha_{\rm L}}{\hat{N}H}\fr{\D}{\D t}\bra{\fr{\beta_1}{1+\alpha_{\rm L}}}}, \qquad
    c_2=\fr{2(1+\alpha_{\rm L})\gamma_1}{\MD}, \\
    &c_3=-4(1+\alpha_{\rm L}), \qquad
    c_4=2(1+\alpha_{\rm T}), \qquad
    c_5=4(1+\alpha_{\rm H})+\fr{4(1+\alpha_{\rm T})\beta_1}{1+\alpha_{\rm L}}, \\
    &c_6=H^2\bra{\alpha_{\rm K}+\fr{6\alpha_{\rm B}^2}{1+\alpha_{\rm L}}}-\fr{6}{\hat{N}a^3M^2}\fr{\D}{\D t}\bra{\fr{a^3M^2H\alpha_{\rm B}\beta_1}{1+\alpha_{\rm L}}}-\fr{c_1^2}{24(1+\alpha_{\rm L})}, \qquad
    c_7=-\fr{\gamma_1}{6\MD}c_1, \\
    &c_8=-\fr{\alpha_{\rm L}\gamma_1^2}{6\MD^2}, \qquad
    c_9=\fr{c_1}{3}, \qquad
    c_{10}=\fr{2\alpha_{\rm L}\gamma_1}{3\MD}, \qquad
    c_{11}=-\fr{2}{3}\alpha_{\rm L}.
    \end{split}
    \ee
In the above action, there is no time derivative acting on $\dn$ or $\psi$ in the action, implying that $\dn$ and $\psi$ can be removed by use of their EOMs.
After this manipulation, we have
    \begin{align}
    \delta_2S_S
    =\int \D t\D^3x\,\hat{N}a^3\fr{M^2}{2}
    \brb{C_1\bra{\fr{\dot{\ti{\zeta}}}{\hat{N}}}^2+\ti{\zeta}\bra{C_2-C_3\fr{\Delta}{a^2}}\fr{\Delta\ti{\zeta}}{a^2}},
    \label{qac_scalar_master}
    \end{align}
with
    \be
    C_1=\fr{6[c_1^2+24c_6(1+\alpha_{\rm L})]}{c_1^2+24c_6\alpha_{\rm L}}, \qquad
    C_2=-c_4+\fr{6}{\hat{N}aM^2}\fr{\D}{\D t}\bra{\fr{aM^2c_1c_5}{c_1^2+24c_6\alpha_{\rm L}}}, \qquad
    C_3=\fr{6c_5^2\alpha_{\rm L}}{c_1^2+24c_6\alpha_{\rm L}}.
    \ee
Then, the conditions for the absence of ghost/gradient instabilities can be read off as
    \be
    C_1>0, \qquad
    C_2>0, \qquad
    C_3 \ge 0.
    \ee
Note that a term with higher spatial derivatives could show up when $\alpha_{\rm L}\ne 0$.
As a result, the dispersion relation of the scalar perturbation has a term of fourth order in the wavenumber.
This is related to the scordatura effect~\cite{Motohashi:2019ymr}, which is implemented by default in generic U-DHOST theories~\cite{DeFelice:2022xvq}.
In the case of GDH theories, as it should, such a higher-derivative term vanishes and we have an ordinary quadratic dispersion relation as in Horndeski theories.

Let us now introduce a matter field.
For demonstration purposes, we consider a k-essence scalar field~\cite{ArmendarizPicon:1999rj} described by the following action as a choice of the matter field:
    \be
    S_{\rm m}[g,\sigma]=\int \D^4x\sqrt{-g}\,P(X_\sigma), \qquad
    X_\sigma\coloneqq g^\mn\pa_\mu\sigma\pa_\nu\sigma,
    \label{Sm}
    \ee
with $\sigma$ denoting the matter scalar field which is minimally coupled to gravity.
We assume that $\sigma$ has a spatially homogeneous background~$\hat{\sigma}(t)$, for which the background value of $X_\sigma$ is given by $\hat{X}_\sigma=-\dot{\hat{\sigma}}^2/\hat{N}^2$.
The background energy density and pressure of $\sigma$ are respectively written as
    \be
    \hat{\rho}_{\rm m}=2\hat{X}_\sigma P'(\hat{X}_\sigma)-P(\hat{X}_\sigma), \qquad
    \hat{p}_{\rm m}=P(\hat{X}_\sigma),
    \ee
where a prime denotes the derivative of $P$ with respect to its argument.
As mentioned earlier and discussed in detail in Appendix~\ref{AppB}, when a matter field is present, one has to take into account its effect on the background EOMs.
Practically, this can be achieved by incorporating the quadratic action~\eqref{S_0_GDH_quad} on top of $\delta_2S_{\rm g}$ in \eqref{EFT_GDT} and the quadratic part of the matter action~\eqref{Sm}.
[See Eq.~\eqref{S_total}.]
Then, it is straightforward to study scalar perturbations on a homogeneous and isotropic background.
In doing so, the matter scalar field~$\sigma$ is assumed to be of the form~$\sigma=\hat{\sigma}(t)+\delta\sigma$, where $\delta\sigma$ denotes the perturbation.
As in the vacuum case, after redefining $\zeta$ as in \eqref{tilde_zeta}, one can integrate out $\nu$ and $\psi$ to obtain the following action:
    \begin{align}
    \delta_2S_S
    =\int \D t\D^3x\,\hat{N}a^3M^2
    \sum_{A,B=1}^2\brb{\fr{1}{2\hat{N}^2}\dot{v}^A\mK_{AB}(t,\Delta)\dot{v}^B+\fr{1}{\hat{N}}\dot{v}^A\mM_{AB}(t,\Delta)v^B-\fr{1}{2}v^A\mW_{AB}(t,\Delta)v^B},
    \end{align}
where we have imposed the set of conditions~\eqref{EFT-param_relation}.
Here, $v^A=(\ti{\zeta},\delta\sigma/\hat{\sigma})$ and $\mK$, $\mM$, and $\mW$ are operator-valued matrices.
Without loss of generality, one can choose $\mK$ and $\mW$ to be symmetric and $\mM$ to be antisymmetric.
In the vacuum limit, we recover the quadratic action~\eqref{qac_scalar_master}.
In the case of $\alpha_{\rm L}=0$ (which is the case for GDH theories), the $\Delta$-dependence is given by
    \be
    \mK_{AB}=\mK^{(0)}_{AB}-\mK^{(2)}_{AB}\fr{\Delta}{a^2}, \qquad
    \mM_{AB}=\mM^{(0)}_{AB}-\mM^{(2)}_{AB}\fr{\Delta}{a^2}, \qquad
    \mW_{AB}=\mW^{(0)}_{AB}-\mW^{(2)}_{AB}\fr{\Delta}{a^2}, \qquad
    \ee
where the coefficient matrices on the right-hand sides are just functions of $t$.\footnote{For $\alpha_{\rm L}\ne 0$, a derivative operator of the form~$F(t)+\alpha_{\rm L}(\Delta/a^2)$ shows up in the denominator, with $F(t)$ being some function of $t$. In other words, the Lagrangian acquires nonlocal interactions.}
In particular, we have
    \be
    \mK^{(2)}_{AB}=-\fr{1728\hat{p}_{\rm m}\hat{X}^2\hat{f}_{0\mZ}}{M^2c_1^2\hat{f}_0}\begin{pmatrix}1&0\\0&0\end{pmatrix}, \qquad
    \mM^{(2)}_{AB}=\fr{3(\hat{\rho}_{\rm m}+\hat{p}_{\rm m})\hat{N}\hat{\sigma}}{M^2c_1\dot{\hat{\sigma}}}\bra{\fr{\gamma_1}{\MD}+\fr{144\hat{p}_{\rm m}\hat{X}^2\hat{f}_{0\mZ}}{M^2c_1\hat{f}_0}}\begin{pmatrix}0&-1\\1&0\end{pmatrix}. \qquad
    \ee
If these matrices are nonvanishing, then one has higher-order spatial derivatives in the EOMs.
This implies that matter-coupled GDH theories would have a shadowy mode, which is consistent with the result of \cite{Ikeda:2023ntu}.
On the other hand, if we impose $\gamma_1=0$ and $f_{0\mZ}=0$ (which is the case for DH theories), there are no higher-order derivatives and we recover the result of \cite{Langlois:2017mxy}.

\section{Conclusions}\label{sec:conc}

The higher-derivative generalization of invertible disformal transformations constructed in Ref.~\cite{Takahashi:2021ttd} offers a possibility to generate more general ghost-free scalar-tensor theories than ever.
Starting with the Horndeski class, one obtains the GDH class~\cite{Takahashi:2022mew}, which form the most general class of ghost-free scalar-tensor theories to this date.
Moreover, one could go further by starting with the U-DHOST class at the price of introducing a shadowy mode, and we call this class the GDU class.
In the present paper, we have extended the EFT of cosmological perturbations to accommodate GDH theories as well as GDU theories.
In doing so, we have restricted ourselves to the transformation of the form~\eqref{GDT_consistent}, which allows for consistent matter coupling without Ostrogradsky ghosts~\cite{Takahashi:2022mew,Naruko:2022vuh,Takahashi:2022ctx}.
The extended EFT involves three new parameters~$\gamma_1$, $\gamma_2$, and $\gamma_3$, on top of those that already exist in the EFT for DHOST theories [see Eq.~\eqref{EFT_GDT}].
The new parameters multiply higher spatial derivatives of the perturbation of the lapse function, which originate from the new disformal factor~$f_2$ introduced in the generalized disformal transformation~\eqref{GDT_consistent}.
Based on our extended EFT, we have studied scalar and tensor cosmological perturbations.
We have obtained the condition under which GWs propagate at the speed of light, to be consistent with the nearly simultaneous detection of the GW event~GW170817 and the gamma-ray burst~170817A emitted from a binary neutron star merger.
Also, we have derived the quadratic action for the scalar perturbations to see that the dispersion relation in general involves terms with higher-order spatial derivatives, which would be related to the existence of a shadowy mode.
In the vacuum case, this effect is controlled by the parameter~$\alpha_{\rm L}$, which can be nonvanishing in U-DHOST and GDU theories.
In the case of GDH theories where $\alpha_{\rm L}=0$, the vacuum dispersion relation has the ordinary quadratic form.
On the other hand, if a matter field is present, higher-order spatial derivatives show up even in GDH theories.
If we restrict ourselves to DH theories, the action involves only up to second-order derivatives, which is consistent with the result of Ref.~\cite{Langlois:2017mxy}.

A possible immediate application of our extended EFT would be the study of the CMB.
Indeed, the authors of Refs.~\cite{Hiramatsu:2020fcd,Hiramatsu:2022fgn} showed that the parameter~$\beta_1$, which is peculiar to DHOST theories, affects the angular power spectrum of the temperature anisotropies through the integrated Sachs-Wolfe effect.
Likewise, it would be intriguing to study how the EFT parameters that are peculiar to GDH/GDU theories can affect the CMB.
Besides, in the context of the EFT for DHOST theories, it is known that the operators associated with the EFT parameters~$\alpha_{\rm H}$ and $\beta_1$ lead to efficient GW decay into scalar perturbations.
Given that GW interferometers have successfully detected GWs, the decay must be suppressed at least in the late-time Universe, which puts a tight constraint on $\alpha_{\rm H}+2\beta_1$~\cite{Creminelli:2018xsv,Creminelli:2019nok} when the EFT is applied to dark energy.
Also, there are possible ghost/gradient instabilities in the presence of a GW background, which yields another constraint on the parameter~$\alpha_{\rm B}$~\cite{Creminelli:2019kjy}.
Considerations of these kinds might in principle put tight constraints on the parameters of our extended EFT~\cite{GWdecay_prep}.
We leave these issues for future work.


\acknowledgments{
The authors would like to thank Keitaro Tomikawa and Vicharit Yingcharoenrat for useful discussions.
K.T.~was supported by JSPS (Japan Society for the Promotion of Science) KAKENHI Grant Nos.~JP21J00695, JP22KJ1646, and JP23K13101.
M.M.~was supported by the Portuguese national fund through the Funda\c{c}\~{a}o para a Ci\^encia e a Tecnologia in the scope of the framework of the Decree-Law 57/2016 of August 29, changed by Law 57/2017 of July 19, and the Centro de Astrof\'{\i}sica e Gravita\c c\~ao through the Project~No.~UIDB/00099/2020.
H.M.~was supported by JSPS KAKENHI Grant No.~JP22K03639.
}


\appendix

\section{Generalized disformal U-DHOST theories}\label{AppA}

U-DHOST theories were proposed as a possible ghost-free extension of DHOST theories by allowing for the existence of a shadowy mode~\cite{DeFelice:2018mkq,DeFelice:2021hps}.
By exploiting the generalized disformal transformation, we can construct more general theories, i.e., GDU theories.
In this appendix, we argue that the quadratic action~\eqref{EFT_GDT} accommodates GDU theories as well as GDH theories.

For concreteness, we consider GDU theories that are generated from quadratic higher-order scalar-tensor theories whose action is given by
	\be
	S=\int \D^4x\sqrt{-g}\brb{G_2(\phi,X)+G_3(\phi,X)\Box \phi+G_4(\phi,X)R+\sum_{I=1}^{5}A_I(\phi,X)L_I^{(2)}}, \label{HOST}
	\ee
where $G_2$, $G_3$, $G_4$, and $A_I$'s are functions of $(\phi,X)$ and
	\be
	L_1^{(2)}\coloneqq \phi^{\mn}\phi_{\mn}, \quad
	L_2^{(2)}\coloneqq (\Box\phi)^2, \quad
	L_3^{(2)}\coloneqq \phi^\mu\phi_{\mn}\phi^\nu\Box\phi, \quad
	L_4^{(2)}\coloneqq \phi^\mu\phi_{\mn}\phi^{\nu\la}\phi_\la, \quad
	L_5^{(2)}\coloneqq (\phi^\mu\phi_{\mn}\phi^\nu)^2.
	\ee
Under the unitary gauge, the degeneracy condition reads~\cite{Langlois:2015cwa,DeFelice:2018mkq,DeFelice:2021hps}
    \be
    4\brb{X(A_1+3A_2)+2G_4}\brb{A_1+A_2+X(A_3+A_4)+X^2A_5}=3X(2A_2+XA_3+4G_{4X})^2, \label{DCU}
    \ee
which defines the U-DHOST class.
Note that Horndeski theories amount to
    \be
    A_1=-A_2=2G_{4X}, \qquad
    A_3=A_4=A_5=0.
    \ee

In order to discuss the mapping between higher-order scalar-tensor theories and the EFT of cosmological perturbations, it is useful to rewrite the action~\eqref{HOST} in the form
    \begin{align}
    S=\int \D^4x\sqrt{-g}\,\bigg\{&G_2-Xg_{3\phi}-2\sqrt{-X}\bra{Xg_{3X}+G_{4\phi}}K+G_4\Rs+\bra{G_4-XA_1}K^\mu_\nu K^\nu_\mu-\bra{G_4+XA_2}K^2 \nonumber \\
    &+\bra{2A_2+XA_3+4G_{4X}}K\fr{\phi^\mu\na_\mu X}{2\sqrt{-X}}+\brb{A_1+A_2+X(A_3+A_4)+X^2A_5}\fr{(\phi^\mu\na_\mu X)^2}{4X^2} \nonumber \\
    &+(2A_1+XA_4-4G_{4X})\bra{g^\mn-\fr{\phi^\mu\phi^\nu}{X}}\fr{\na_\mu X\na_\nu X}{4X}\bigg\}\,,
    \label{HOST2}
    \end{align}
where the function~$g_3(\phi,X)$ has been defined so that $g_3+2Xg_{3X}=G_3$.
Also, $K_\mn$ and $\Rs$ are the extrinsic curvature and the spatial Ricci scalar associated with the constant-$\phi$ slicing, respectively [see Eq.~\eqref{phi_der_unitary}].
Here, we have assumed $X<0$, i.e., the scalar field has a timelike profile.
Starting with this action, one can move to the frame where
    \be
    A_1+A_2+X(A_3+A_4)+X^2A_5
    =2A_1+XA_4-4G_{4X}
    =0, \label{condition_on_HOST_functions}
    \ee
by performing a conventional disformal transformation of the form~\eqref{conv_disformal} with an appropriate choice of the functions~$f_0$ and $f_1$.
For U-DHOST theories that satisfy \eqref{DCU}, this leads to $2A_2+XA_3+4G_{4X}=0$.
Then, the terms with higher derivatives in \eqref{HOST2} [i.e., those in the second and third lines] are removed, and we are left with the following action:
    \be
    S=\int \D^4x\sqrt{-g}\brb{G_2-Xg_{3\phi}-2\sqrt{-X}\bra{Xg_{3X}+G_{4\phi}}K+G_4\Rs+\bra{G_4-XA_1}K^\mu_\nu K^\nu_\mu-\bra{G_4+XA_2}K^2}.
    \label{U-DHOST_seed}
    \ee
It should be noted that any (generalized) disformal transformation of quadratic U-DHOST theories can be generated from \eqref{U-DHOST_seed} by use of some (generalized) disformal transformation.
This is because the functional composition of two (generalized) disformal transformations is again a (generalized) disformal transformation~\cite{Takahashi:2021ttd}.

For U-DHOST theories described by the action~\eqref{U-DHOST_seed}, one can follow the procedure of \S\ref{ssec:seed_action} to obtain the quadratic action for cosmological perturbations, which has the form of \eqref{EFT_lowest}.
In particular, we have
    \be
    \ti{M}^2=2(G_4-XA_1), \qquad
    \aH_{\rm L}=\fr{3}{2}\fr{X(A_1+A_2)}{G_4-XA_1}, \qquad
    \aH_{\rm T}=\fr{XA_1}{G_4-XA_1}, \qquad
    \aH_{\rm H}=\fr{X(A_1-2G_{4X})}{G_4-XA_1}.
    \label{alpha_param_U-DHOST_seed}
    \ee
We note that $\aH_{\rm L}$ and $\aH_{\rm H}$ can be nonvanishing, which is in sharp contrast to the case of Horndeski theories [see Eq.~\eqref{alpha_param_Horndeski}].
Therefore, \eqref{EFT_GDT} accommodates GDU theories.

\section{Coupling with matter fields}\label{AppB}

In deriving the quadratic action for cosmological perturbations in \S\ref{sec:EFT}, we did not explicitly consider matter fields for simplicity.
However, if one would like to use the EFT as a model of the late-time Universe, matter fields have to be taken into account.
In this appendix, we discuss how the existence of matter fields affects the EFT action.

In \S\ref{ssec:seed_action}, we removed the following part from the action by use of vacuum background EOMs [see Eq.~\eqref{S_0}]:
    \be
    S_0=\int \D^4x\sqrt{-g}\brb{\mE_0+\mE_1\bra{\fr{\delta N}{\hat{N}}-\fr{\delta N^2}{\hat{N}^2}}}, \label{S_0_App}
    \ee
where $\mE_0$ and $\mE_1$ are defined by \eqref{BGEOMs}.
Written explicitly in terms of the coefficient functions~$G_I$ in \eqref{S_H},
    \begin{align}
    \mE_0&=G_2-XG_{3\phi}+\fr{\dot{\phi}\dot{X}}{N^2}G_{3X}+2\bra{3H^2+\fr{2\dot{H}}{N}}\bra{G_4-2XG_{4X}}-4H\fr{\dot{X}}{N}\bra{G_{4X}+2XG_{4XX}} \nonumber \\
    &\quad +\fr{\dot{\phi}}{NX}\bra{4HX+\fr{\dot{X}}{N}}G_{4\phi}-2\fr{\dot{\phi}}{N}\bra{4HX-\fr{\dot{X}}{N}}G_{4\phi X}-2XG_{4\phi\phi}, \label{E0E1} \\
    \mE_1&=G_2-2XG_{2X}+XG_{3\phi}+6HX\fr{\dot{\phi}}{N}G_{3X}+6H^2\bra{G_4-4XG_{4X}-4X^2G_{4XX}} +6H\fr{\dot{\phi}}{N}\bra{G_{4\phi}+2XG_{4\phi X}}-\mE_0. \nonumber
    \end{align}
Then, in \S\ref{ssec:GDH_EFT}, we obtained the EFT action that accommodates GDH theories by performing the generalized disformal transformation on the seed EFT action with the part~\eqref{S_0_App} subtracted.
When matter fields are introduced, the results of \S\ref{ssec:GDH_EFT} for the quadratic action of the gravity sector~$\delta_2S_{\rm g}$ remain correct, but one can no longer simply remove the action~\eqref{S_0_App} since the background EOMs are affected by the matter fields.
One has to study how the action~\eqref{S_0_App} is transformed under the generalized disformal transformation to correctly incorporate the effect of the matter fields.

Performing the generalized disformal transformation, the action~\eqref{S_0_App} is transformed as
    \begin{align}
    S_0&\mapsto \int \D^4x\sqrt{-\mybar{g}}\brb{\mybar{\mE}_0+\mybar{\mE}_1\bra{\fr{\delta \mybar{N}}{\hat{\mybar{N}}}-\fr{\delta \mybar{N}^2}{\hat{\mybar{N}}^2}}} \nonumber \\
    &=\int \D^4x\sqrt{-g}\left\{\vphantom{\fr{\hat{X}^2\hat{\mybar{X}}_X^2}{2\hat{\mybar{X}}^2}}
    \hat{\mJ}\mybar{\mE}_0+\bra{\fr{\hat{X}\hat{\mybar{X}}_X}{\hat{\mybar{X}}}\hat{\mJ}\mybar{\mE}_1-2\hat{X}\hat{\mJ}_X\mybar{\mE}_0}\fr{\delta N}{\hat{N}}
    +\fr{6\hat{X}^2\hat{f}_{0\mZ}}{\hat{f}_0}\hat{\mJ}\mybar{\mE}_0\fr{(\pa_i\delta N)^2}{\hat{N}^2a^2}\right. \nonumber \\
    &\quad \left.+\brb{\mybar{\mE}_0\bra{3\hat{X}\hat{\mJ}_X+2\hat{X}^2\hat{\mJ}_{XX}}
    -\hat{\mJ}\mybar{\mE}_1\bra{\fr{3\hat{X}\hat{\mybar{X}}_X}{2\hat{\mybar{X}}}-\fr{\hat{X}^2\hat{\mybar{X}}_X^2}{2\hat{\mybar{X}}^2}+\fr{\hat{X}^2\hat{\mybar{X}}_{XX}}{\hat{\mybar{X}}}+\fr{2\hat{X}^2\hat{\mybar{X}}_X\hat{\mJ}_X}{\hat{\mybar{X}}\hat{\mJ}}}}\fr{\delta N^2}{\hat{N}^2}\right\}+\mO(\delta^3), \label{S_0_GDH}
    \end{align}
where $\mybar{\mE}_0$ and $\mybar{\mE}_1$ are the barred counterparts of $\mE_0$ and $\mE_1$ in \eqref{E0E1}, respectively, and we have defined $\mJ=f_0\mF^{1/2}$.
We now introduce matter fields that are minimally coupled to GDH gravity.
Then, the background EOMs are given by
    \be
    \hat{\mJ}\mybar{\mE}_0=-\hat{p}_{\rm m}, \qquad
    \fr{\hat{X}\hat{\mybar{X}}_X}{\hat{\mybar{X}}}\hat{\mJ}\mybar{\mE}_1-2\hat{X}\hat{\mJ}_X\mybar{\mE}_0=\hat{\rho}_{\rm m}+\hat{p}_{\rm m},
    \ee
where $\hat{\rho}_{\rm m}$ and $\hat{p}_{\rm m}$ are the background energy density and pressure of the 
matter fields, respectively.
These equations yield $\mybar{\mE}_0$ and $\mybar{\mE}_1$ in terms of $\hat{\rho}_{\rm m}$ and $\hat{p}_{\rm m}$ as
    \be
    \mybar{\mE}_0=-\fr{\hat{p}_{\rm m}}{\hat{\mJ}}, \qquad
    \mybar{\mE}_1=\fr{\hat{\mybar{X}}}{\hat{X}\hat{\mybar{X}}_X\hat{\mJ}}\bra{\hat{\rho}_{\rm m}+\hat{p}_{\rm m}-2\hat{p}_{\rm m}\fr{\hat{X}\hat{\mJ}_X}{\hat{\mJ}}}.
    \ee
Substituting these expressions back into \eqref{S_0_GDH}, we see that the contribution to the quadratic action is given by
    \begin{align}
    \delta_2S_0=
    \delta_2\int \D^4x\,\hat{N}\sqrt{\gamma}\Bigg[\!&-\hat{p}_{\rm m}+\hat{\rho}_{\rm m}\fr{\delta N}{\hat{N}}
    -\hat{p}_{\rm m}\hat{X}^2\bra{\fr{\hat{\mybar{X}}_X\hat{\mJ}_X}{\hat{\mybar{X}}\hat{\mJ}}-\fr{2\hat{\mybar{X}}_{XX}\hat{\mJ}_X}{\hat{\mybar{X}}_X\hat{\mJ}}-\fr{4\hat{\mJ}_X^2}{\hat{\mJ}^2}-\fr{2\hat{\mJ}_{XX}}{\hat{\mJ}}}\fr{\delta N^2}{\hat{N}^2} \nonumber \\
    &-\fr{\hat{\rho}_{\rm m}+\hat{p}_{\rm m}}{2}\bra{1-\fr{\hat{X}\hat{\mybar{X}}_X}{\hat{\mybar{X}}}+\fr{2\hat{X}\hat{\mybar{X}}_{XX}}{\hat{\mybar{X}}_X}+\fr{4\hat{X}\hat{\mJ}_X}{\hat{\mJ}}}\fr{\delta N^2}{\hat{N}^2}
    -\fr{6\hat{p}_{\rm m}\hat{X}^2\hat{f}_{0\mZ}}{\hat{f}_0}\fr{(\pa_i\delta N)^2}{\hat{N}^2a^2}\Bigg]. \label{S_0_GDH_quad}
    \end{align}
Note that the last term is nonvanishing only when $f_0$ has a nontrivial dependence on $\mZ$.
In order to study the dynamics of cosmological perturbations in the presence of matter fields, one has to take this contribution into account on top of the matter action.
More concretely, the total quadratic action is written as
    \be
    \delta_2S=\delta_2S_{\rm g}+\delta_2S_0+\delta_2S_{\rm m}.
    \label{S_total}
    \ee
Here, $\delta_2S_{\rm g}$ is given by \eqref{EFT_GDT}, $\delta_2S_0$ is the one obtained above in \eqref{S_0_GDH_quad}, and $\delta_2S_{\rm m}$ denotes the contribution from the action~$S_{\rm m}$ of (minimally coupled) matter fields.


\bibliographystyle{mybibstyle}
\bibliography{bib}

\clearpage

\thispagestyle{empty}

\pdfbookmark[0]
    {Erratum}{erratum-title}

\begin{center}
    {\large\bfseries
    Erratum: Effective description of generalized disformal theories
    }

    \vspace{1.5em}

    Kazufumi Takahashi,\textsuperscript{1}
    Masato Minamitsuji,\textsuperscript{2}
    and Hayato Motohashi\textsuperscript{3}

    \vspace{0.5em}

    \begin{minipage}{0.9\textwidth}
    \begin{center}
        \small\it
        \textsuperscript{\rm 1}Research Center for the Early Universe (RESCEU), Graduate School of Science, The University of Tokyo, Hongo 7-3-1, Bunkyo-ku, Tokyo 113-0033, Japan\\
        \textsuperscript{\rm 2}Faculty of Health Sciences, Butsuryo College of Osaka, Sakai 593-8328, Osaka, Japan\\
        \textsuperscript{\rm 3}Department of Physics, Tokyo Metropolitan University, 1-1 Minami-Osawa, Hachioji, Tokyo 192-0397, Japan
    \end{center}
    \end{minipage}

    \vspace{1em}
\end{center}

In Appendix~A of the original article, we argued that the discussion in the main text also applies to theories generated from quadratic U-DHOST theories.
We have found that this statement is not valid in full generality.
As we clarify below, the discussion applies only to a subclass of such theories.

Let us consider a conventional disformal transformation,
    \begin{align*}
    g_{\mu\nu}
    \to \Omega(\phi,X)g_{\mu\nu}+\Gamma(\phi,X)\phi_\mu\phi_\nu,
    \end{align*}
where $\phi_\mu\coloneqq\partial_\mu\phi$, and define
    \begin{align*}
    \bar{X}(\phi,X)
    \coloneqq\frac{X}{\Omega(\phi,X)+X\Gamma(\phi,X)}.
    \end{align*}
We also introduce
    \begin{align*}
    \omega(\phi,\bar{X})
    \coloneqq\ln\Omega\left(\phi,X(\phi,\bar{X})\right),
    \end{align*}
where $X(\phi,\bar{X})$ denotes the local inverse of $\bar{X}(\phi,X)$.
Under the disformal transformation, the coefficients of the last two terms in Eq.~\eqref{HOST2} are transformed into
    \begin{align*}
    &A_1+A_2+\bar{X}(A_3+A_4)+\bar{X}^2A_5+3(2A_2+\bar{X}A_3+4\partial_{\bar{X}}G_{4})\bar{X}\partial_{\bar{X}}\omega+3\left(A_1+3A_2+\frac{2G_4}{\bar{X}}\right)\left(\bar{X}\partial_{\bar{X}}\omega\right)^2, \\
    &2A_1+\bar{X}A_4-4\partial_{\bar{X}}G_{4}-4\left(\frac{G_{4}}{\bar{X}}-2\partial_{\bar{X}}G_{4}\right)\bar{X}\partial_{\bar{X}}\omega+2\bar{X}G_{4}\left(\partial_{\bar{X}}\omega\right)^2,
    \end{align*}
up to nonvanishing overall factors.
Here, the functions~$G_4$ and $A_I$'s are evaluated at $(\phi,\bar{X})$.

In the original article, we argued that the conformal factor~$\Omega$ and the disformal factor~$\Gamma$ could be chosen so that these two coefficients vanish.
However, in terms of $\bar{X}$, the two coefficients are both controlled by the single function~$\omega(\phi,\bar{X})$.
The disformal factor~$\Gamma$ determines the relation between $X$ and $\bar{X}$, but does not provide an independent functional freedom for the two coefficients to vanish simultaneously.
Equivalently, the two conditions constitute two quadratic equations for $\partial_{\bar{X}}\omega$, which can be satisfied simultaneously only when the two quadratic equations possess a common real root.

It should be noted that, upon using the degeneracy condition~\eqref{DCU} for quadratic U-DHOST theories, the first of the above two coefficients becomes a complete square.
The condition that this coefficient vanish uniquely fixes
    \begin{align*}
    \partial_{\bar{X}}\omega
    =-\frac{2A_2+\bar{X}A_3+4\partial_{\bar{X}}G_{4}}{2\left[2G_{4}+\bar{X}(A_1+3A_2)\right]},
    \end{align*}
provided that $2G_{4}+\bar{X}(A_1+3A_2)\ne 0$.
For this choice of $\partial_{\bar{X}}\omega$, the coefficient of the term proportional to $K\phi^\mu\nabla_\mu X$ in Eq.~\eqref{HOST2} vanishes automatically, whereas the coefficient of the last term vanishes only if the expression above also satisfies the condition obtained by setting the second of the above two coefficients to zero.

Therefore, the results of the original article apply only to a subclass of quadratic U-DHOST theories.
Thus, Eq.~\eqref{condition_on_HOST_functions} may consistently be imposed as a defining condition of the subclass considered in the original article, but it cannot be reached from a generic quadratic U-DHOST theory by an invertible conventional disformal transformation.
A treatment of generic quadratic U-DHOST theories would require retaining the last term of Eq.~\eqref{HOST2} in the seed action given in Eq.~\eqref{U-DHOST_seed}, which is beyond the scope of this erratum.

All subsequent calculations and results remain valid within this subclass; the correction affects only the claimed range of applicability of the analysis.


\acknowledgments{
The authors would like to thank Tsutomu Kobayashi for pointing out the issue addressed in this erratum and for helpful discussions.
}

\end{document}